\documentclass[pra,twocolumn]{revtex4-1}

\usepackage{graphicx}
\usepackage{array}
\usepackage{bm}
\usepackage[english]{babel}
\usepackage{dcolumn}
\usepackage{amssymb}
\usepackage{amsmath}

\begin{document}

\title{A general comparison theorem}

\author{Claude \surname{Semay}}
\email[E-mail: ]{claude.semay@umons.ac.be}
\affiliation{Service de Physique Nucl\'{e}aire et Subnucl\'{e}aire,
Universit\'{e} de Mons,
Acad\'{e}mie universitaire Wallonie-Bruxelles,
Place du Parc 20, 7000 Mons, Belgium}
\date{\today}

\begin{abstract}
Using the Hellmann-Feynman theorem, a general comparison theorem is established for an eigenvalue equation of the form $(T+V)|\psi\rangle = E|\psi\rangle$, where $T$ is a kinetic part which depends only on momentums and $V$ is a potential which depends only on positions. We assume that $H^{(1)}=T+V^{(1)}$ and $H^{(2)}=T+V^{(2)}$ ($H^{(1)}=T^{(1)}+V$ and $H^{(2)}=T^{(2)}+V$) support both discrete eigenvalues $E^{(1)}_{\{\alpha\}}$ and $E^{(2)}_{\{\alpha\}}$, where ${\{\alpha\}}$ represents a set of quantum numbers. We prove that, if $V^{(1)} \le V^{(2)}$ ($T^{(1)} \le T^{(2)}$) for all position (momentum) variables, then the corresponding eigenvalues are ordered $E^{(1)}_{\{\alpha\}} \le E^{(2)}_{\{\alpha\}}$. Some analytical applications are given.
\end{abstract}



\maketitle

The comparison theorem of quantum mechanics states that, for some eigenvalue equations, if two real potentials are ordered, $V^{(1)} \le V^{(2)}$, then each corresponding pair of eigenvalues is ordered $E^{(1)}_{\{\alpha\}} \le E^{(2)}_{\{\alpha\}}$ (${\{\alpha\}}$ represents a set of quantum numbers). This can be shown for Hamiltonians which are bounded from below by using the Ritz variational principle. But, such a procedure is not applicable for the corresponding Dirac problem since the Dirac Hamiltonian is not bounded from below. Nevertheless, the comparison theorem has also been proved for a Dirac equation with a potential monotone in a parameter \cite{Hall99,Hall08,Hall10}.

Using the Hellmann-Feynman theorem \cite{Hell35}, we shall see that the comparison theorem can be applied to a great class of eigenvalue problems written in the form 
\begin{equation}
\label{TpV}
(T+V)|\psi\rangle = E|\psi\rangle, 
\end{equation}
where $T$ is a kinetic part which depends only on momentums and $V$ is a potential which depends only on positions. No assumption is made about the number of particles, and it is not necessary that the Hamiltonian is bounded from below. We have not seen this presentation elsewhere, although it is related to ideas presented in \cite{Hall10}. As the Klein-Gordon equation is not of the form (\ref{TpV}), the comparison theorem presented here does not apply, but results about the ordering of the spectra can been obtained \cite{Hall08b,Hall10}. 

The Hellmann-Feynman theorem states that if the Hamiltonian of a system is $H(a)$ where $a$ is a parameter, and that the eigenvalue equation for a bound state is 
\begin{equation}
\label{HpsiEpsi}
H(a)|a\rangle = E(a)|a\rangle,
\end{equation}
where $E(a)$ is the energy and $|a\rangle$ the normalized associated eigenstate, then
\begin{equation}
\label{HFtheor}
\frac{\partial E(a)}{\partial a} = \left\langle a \left|\frac{\partial H(a)}{\partial a}\right|a\right\rangle.
\end{equation}

We consider two Hamiltonians $H^{(1)}$ and $H^{(2)}$ such that
\begin{equation}
\label{H2gtH1}
\langle \phi | H^{(2)} - H^{(1)} | \phi \rangle \ge 0, \quad \forall\ | \phi \rangle.
\end{equation}
Let us assume that the Hamiltonian 
\begin{equation}
\label{HaV}
H(a)=(1-a) H^{(1)} + a H^{(2)}
\end{equation}
possesses a number (finite or infinite) of well defined eigenvalues $E_{\{\alpha\}}(a)$ characterized by a set of quantum numbers ${\{\alpha\}}$, for $0 \le a \le 1$ \cite{fn1}. If $\left|a;{\{\alpha\}}\right\rangle$ is the corresponding eigenstate, the Hellmann-Feynman theorem directly yields
\begin{equation}
\label{HFV1V2}
\frac{\partial E_{\{\alpha\}}(a)}{\partial a} = \left\langle a;{\{\alpha\}} \left| H^{(2)} - H^{(1)}\right|a;{\{\alpha\}}\right\rangle.
\end{equation}
It is a positive number due to the hypothesis (\ref{H2gtH1}). So, $E_{\{\alpha\}}(a)$ is an increasing function of $a$ and 
\begin{equation}
\label{EcompV1V2}
E_{\{\alpha\}}(0) =E^{(1)}_{\{\alpha\}} \le E^{(2)}_{\{\alpha\}} =E_{\{\alpha\}}(1),
\end{equation}
where $E^{(1)}_{\{\alpha\}}$ and $E^{(2)}_{\{\alpha\}}$ are respectively eigenvalues of Hamiltonians $H^{(1)}$ and $H^{(2)}$.
Condition (\ref{H2gtH1}) is not necessarily easy to verify for arbitrary Hamiltonians. It is then interesting to look at two particular simpler situations.

Let us first consider two Hamiltonians $H^{(1)}=T+V^{(1)}$ and $H^{(2)}=T+V^{(2)}$ such that $V^{(1)} \le V^{(2)}$ for all values of the position variables appearing in these potentials. Relation (\ref{H2gtH1}) is satisfied since the mean value is taken for the positive quantity $V^{(2)} - V^{(1)}$. So, the theorem applies in this case. Strictly speaking, the condition $V^{(1)} \le V^{(2)}$ must not be satisfied everywhere in the position space of the potentials. Indeed, some results can be obtained for two-body nonrelativistic problems in which the graphs of the comparison potentials cross each other in a controlled way \cite{Hall92}.

Even if the variety of kinetic operators is much smaller than for potentials, it is worth comparing the spectra of two Hamiltonians $H^{(1)}=T^{(1)}+V$ and $H^{(2)}=T^{(2)}+V$ such that $T^{(1)} \le T^{(2)}$ for all values of the momentum variables appearing in these kinetic operators. For similar reasons as the ones presented above, the theorem also applies in this case. It is not really surprising that the comparison theorem works for both potential and kinetic operator, because one can indifferently consider a position or an momentum representation for Hamiltonians. 

We now give several analytical illustrations of this theorem for different two-body systems, using in particular the following well known cases:
\begin{eqnarray}
\label{Hoh}
H^{(ho)}&=&\frac{\bm p^2}{2 \mu}+\lambda r^2,\ E^{(ho)}_{n,l}=\sqrt{\frac{2 \lambda}{\mu}}Q^{(ho)}_{n,l},\nonumber \\
&&\quad Q^{(ho)}_{n,l}=2 n+l+3/2; \\
\label{Hc}
H^{(c)}&=&\frac{\bm p^2}{2 \mu}-\frac{\kappa}{r},\ E^{(c)}_{n,l}=-\frac{\mu \kappa^2}{2 \left(Q^{(c)}_{n,l}\right)^2},\nonumber \\
&&\quad Q^{(c)}_{n,l}=n+l+1.
\end{eqnarray}

If the potential part of a Hamiltonian possessing a discrete spectra is of the form $g v(r)$ where $v(r)$ is a positive function and $g$ a coupling constant (positive or negative, according to the structure of $v(r)$), the theorem states that the eigenvalues increases with $g$. This is a well known result which can be immediately checked with (\ref{Hoh}) and (\ref{Hc}).

With $T=\bm p^2/(2 \mu)$, let us now consider
\begin{eqnarray}
\label{V1coul}
V^{(1)}(r)&=&-\frac{\kappa}{r}, \\
\label{V2harm}
V^{(2)}(r)&=&\frac{\kappa}{2 r_0^3}r^2-\frac{3\kappa}{2 r_0},
\end{eqnarray}
where $r_0$ is an arbitrary positive distance. These potentials are tangent at $r_0$, and the quantity
\begin{equation}
\label{V2mV1}
V^{(2)}(r)-V^{(1)}(r) = \frac{\kappa}{2 r}\left( \frac{r}{r_0}-1 \right)^2 \left( \frac{r}{r_0}+2 \right)
\end{equation}
is non-negative for all physical values of $r$. Using (\ref{Hoh}) and (\ref{Hc}), the difference between two corresponding eigenvalues is given by
\begin{equation}
\label{E2mE1}
E^{(2)}_{n,l} - E^{(1)}_{n,l}= \frac{\sqrt{\kappa} \left[ x^3 - 3 \left(Q^{(c)}_{n,l}\right)^2 x + 2 Q^{(ho)}_{n,l} \left(Q^{(c)}_{n,l}\right)^2 \right]}{2 r_0 \sqrt{\mu r_0}\left(Q^{(c)}_{n,l}\right)^2},
\end{equation}
where $x=\sqrt{\kappa \mu r_0}$ is an arbitrary positive quantity. It is easy to check that, for positive values of $x$, the polynomial between brackets has one minimum in $x=Q^{(c)}_{n,l}$, with a value equal to $2 \left(Q^{(c)}_{n,l}\right)^2 \left( Q^{(ho)}_{n,l}- Q^{(c)}_{n,l} \right) >0$. So, we have $E^{(2)}_{n,l} - E^{(1)}_{n,l}>0$, in accordance with the comparison theorem.

Provided the potential $V(r)$ does not depend on the mass $m$, the theorem predicts that the eigenvalues of the Hamiltonian $\bm p^2/m+V(r)$ decrease for an increasing mass $m$, while the eigenvalues of the spinless Salpeter Hamiltonian $2\sqrt{\bm p^2+m^2}+V(r)$ increase with the mass $m$. This already known result \cite{Lich89} can be immediately checked on (\ref{Hoh}) and (\ref{Hc}) for the nonrelativistic case. 

Let us now consider the two kinetic operators $T^{(1)}=2\sqrt{\bm p^2+m^2}$ and $T^{(2)}=2m+\bm p^2/m$ which is the nonrelativistic limit of $T^{(1)}$. We have $T^{(1)} \le T^{(2)}$ since $\left(T^{(2)}\right)^2 - \left(T^{(1)}\right)^2=\bm p^4/m^2 \ge 0$. We can conclude that the replacement of the semirelativistic kinetic part in a Hamiltonian by its nonrelativistic counterpart implies an increase of the spectra \cite{Luch96}. Let us consider an explicit example. The eigenvalues $E^{(2)}_{n,l}$ of $H^{(2)}=2m+\bm p^2/m-\kappa/r$ are given by (\ref{Hc}), while only upper bounds $\tilde E^{(1)}_{n,l}$ of the exact eigenvalues $E^{(1)}_{n,l}$ of $H^{(1)}=2\sqrt{\bm p^2+m^2}-\kappa/r$ can be obtained (the value of $\kappa$ is assumed to be low enough to allow the existence of bound states) \cite{Silv09}:
\begin{eqnarray}
\label{E1SR}  
\tilde E^{(1)}_{n,l} &=& 2m\sqrt{1-\frac{\kappa^2}{4 \left(Q^{(c)}_{n,l}\right)^2}}, \\
\label{E2SR}
E^{(2)}_{n,l}&=& 2m-\frac{m \kappa^2}{4 \left(Q^{(c)}_{n,l}\right)^2} .
\end{eqnarray}
Since $\left( E^{(2)}_{n,l} \right)^2 - \left( \tilde E^{(1)}_{n,l} \right)^2 = m^2\kappa^4/\left(16 \left(Q^{(c)}_{n,l}\right)^4\right)>0$, we have $E^{(2)}_{n,l} > \tilde E^{(1)}_{n,l} \ge E^{(1)}_{n,l}$. 

\begin{acknowledgments}
The author thanks the F.R.S.-FNRS for financial support. He also thanks F. Buisseret for a careful reading of the manuscript. 
\end{acknowledgments}


\begin{thebibliography}{99}
\bibitem{fn1} For instance, the Yukawa interaction, which has only a finite number of bound states, is everywhere bounded from above by a harmonic potential, which is characterized by an infinite number of discrete eigenvalues.
\end{thebibliography}
\end{document}